\documentclass[aps,prb,superscriptaddress,twocolumn,showpacs]{revtex4}
\usepackage{graphicx,color}
\usepackage{amsmath}

\begin{document}

%\preprint{\underline{$\rightarrow$ PRB Rapid Comm. subm 1}}

\title{Why could Electron Spin Resonance be observed in a heavy fermion Kondo lattice?}

\author{B.I. Kochelaev}
%\email{Boris.Kochelaev@ksu.ru}
\affiliation{Theoretical Physics Departement, Kazan State University, 420008 Kazan, Russia}

\author{S.I. Belov}
\affiliation{Theoretical Physics Departement, Kazan State University, 420008 Kazan, Russia}

\author{A.M. Skvortsova}
\affiliation{Theoretical Physics Departement, Kazan State University, 420008 Kazan, Russia}

\author{A.S. Kutuzov}
\affiliation{Theoretical Physics Departement, Kazan State University, 420008 Kazan, Russia}

\author{J.~Sichelschmidt}
%\email{Sichelschmidt@cpfs.mpg.de}
\affiliation{Max Planck Institute for Chemical Physics of Solids, 01187 Dresden, Germany}

\author{J. Wykhoff}
\affiliation{Max Planck Institute for Chemical Physics of Solids, 01187 Dresden, Germany}

\author{C. Geibel}
\affiliation{Max Planck Institute for Chemical Physics of Solids, 01187 Dresden, Germany}

\author{F. Steglich}
\affiliation{Max Planck Institute for Chemical Physics of Solids, 01187 Dresden, Germany}

\date{\today}

\begin{abstract}
We develop a theoretical basis for understanding the spin relaxation processes in Kondo lattice systems with heavy fermions as experimentally observed by electron spin resonance (ESR). The Kondo effect leads to a common energy scale that regulates a logarithmic divergence of different spin kinetic coefficients and supports a collective spin motion of the Kondo ions with conduction electrons. We find that the relaxation rate of a collective spin mode is greatly reduced due to a mutual cancelation of all the divergent contributions even in the case of the strongly anisotropic Kondo interaction. The contribution to the ESR linewidth caused by the local magnetic field distribution is subject to motional narrowing supported by ferromagnetic correlations. The developed theoretical model successfully explains the ESR data of YbRh$_{2}$Si$_{2}$ in terms of their dependence on temperature and magnetic field.
\end{abstract}
\maketitle
%
%
%Introduction
%
%
The discovery of electron spin resonance (ESR) in the heavy fermion Kondo lattice YbRh$_{2}$Si$_{2}$ at temperatures below the thermodynamically determined Kondo temperature ($T_{\rm K}= 25$~K) became a great surprise for the condensed matter physics community \cite{sichelschmidt03a}.  According to the common belief, based on the single ion Kondo effect, the ESR signal should not be observable by at least two reasons. Firstly, at temperatures below $T_{\rm K}$ the magnetic moments of the Kondo ions should be screened by the conduction electrons; secondly, the ESR linewidth was expected to be of the order of $T_{\rm K}\equiv 500$~GHz. The experimental results were completely opposite: at X-band (9.4~GHz) and $T=0.7$~K a linewidth of 0.3~GHz was observed and the ESR intensity corresponds to the participation of all Kondo ions with a temperature dependence following a Curie-Weiss law.\cite{wykhoff07b} Moreover, the angular dependence of the resonant magnetic field reflects the tetragonal symmetry of the electric crystal field at the position of the Yb-ion with an extremely anisotropic g-factor ($g_ \bot   = 3.56$, $g_\parallel   = 0.17$ at $T=5$~K). Similar results were obtained later for YbIr$_{2}$Si$_{2}$ \cite{sichelschmidt07a} and the main features of the ESR phenomenon were confirmed recently at very high frequencies up to 360 GHz \cite{schaufuss09a}. Up to now this paradox was not resolved on the basis of the properties of these materials studied by other methods. In particular, the heavy-fermion system YbRh$_{2}$Si$_{2}$ attracted a lot of attention due to the existence of an antiferromagnetic quantum critical point (QCP), at which antiferromagnetic order disappears smoothly as $T\rightarrow 0$, by variations of an external magnetic field, pressure or doping. Near a QCP and up to surprisingly high temperatures, a new phase appears exhibiting non-Fermi liquid  (NFL) behavior. For YbRh$_{2}$Si$_{2}$ the electrical resistivity linearly increases with temperature and the Sommerfeld coefficient of the electronic specific heat diverges logarithmically upon cooling down to 0.3~K. At lower temperatures this divergence becomes even stronger. It seems that many properties of heavy fermions in the NFL state of YbRh$_{2}$Si$_{2}$ can be described in terms of quasi-localized $f$-electrons. In particular, the underlying fluctuations near the QCP can be considered as locally critical, having an atomic length scale \cite{si01a,gegenwart08a}. On the basis of a localized 4$f$ electron approach we successfully studied the temperature dependence of the static magnetic susceptibility of YbRh$_{2}$Si$_{2}$ and YbIr$_{2}$Si$_{2}$ at temperatures below $T_{\rm K}$ \cite{kutuzov08a}.

In this paper we show that the key ingredient of the ESR signal existence in a heavy fermion Kondo lattice in the NFL state is a formation of a collective spin mode of quasi-localized $f$-electrons and wide-
band conduction electrons even in a strongly anisotropic system. We shall discuss, also, the role of the local ferromagnetic fluctuations in YbRh$_{2}$Si$_{2}$ \cite{gegenwart05a} and YbIr$_{2}$Si$_{2}$ \cite{hossain05a}.
%
% Model properties
%

Our basic theoretical model includes the kinetic energy of conduction electrons, the Zeeman energy, the Kondo interaction of the Yb$^{3+}$ ions with conduction electrons, and the coupling between the Yb$^{3+}$ ions via conduction electrons (the RKKY interaction). The lowest multiplet of the free Yb$^{3+}$ ion is $^2$F$_{7/2}$ with total momentum $J = 7/2$. The tetragonal crystal electric field splits this multiplet into four Kramers doublets that are well separated from the lowest (ground) doublet \cite{stockert06a} such that the physics of low energy spin excitations can be described by the lowest doublet. After projection onto the Kramers ground state we obtain an effective Hamiltonian  $\mathcal{H} = \mathcal{H}_0  + \mathcal{H}_{s\sigma }  + \mathcal{H}_{\rm RKKY} $ with
\begin{align}
\mathcal{H}&_0  =  \sum\limits_{{\bf{k}}\alpha } {\varepsilon _{\bf{k}} c_{{\bf{k}}\alpha }^ +  } c_{{\bf{k}}\alpha }  + \nonumber \\
                            & +\sum\limits_j {\left[ {g_\bot  \left( {B_x S_j^x  + B_y S_j^y } \right) + g_\parallel  B_z S_j^z  + g_\sigma  {\boldsymbol{B}}_j }\boldsymbol{\sigma}_j \right]},\label{H0}\\
\mathcal{H}&_{\rm s\sigma }   =  J\sum\limits_j {\left[ {g_ \bot  \left( {S_j^x \sigma _j^x  + S_j^y \sigma _j^y } \right) + g_\parallel  S_j^z \sigma _j^z } \right]},\label{Hssigma} \\ 
\mathcal{H}&_{\rm RKKY}  =  \frac{1}{2}\sum\limits_{ij} {\left[ {I_{ij}^ \bot  \left( {S_i^x S_j^x  + S_i^y S_j^y } \right) + I_{ij}^\parallel  S_i^z S_j^z } \right].} 
\label{HRKKY}
\end{align}
Here $\alpha$ labels the orientation of the conduction electron spin, $\boldsymbol{S}_j$ is the spin-1/2 operator of the $j$-th Yb-ion, $\boldsymbol{B}$ is the external magnetic field multiplied by the Bohr magneton; $g_\parallel, g_\bot$ are the $g$-factors for $\boldsymbol{B}$ aligned parallel and perpendicular to the crystal symmetry axis; $J$ and $I_{ij}$ are the Kondo- and RKKY-coupling constants; $g_{\sigma}$ and $\boldsymbol{\sigma}_j$ denote the $g$-factor and the spin operator of conduction electrons at the position of the Yb-ion. In the following we shall discuss the contributions of the magnetic dipole-dipole and the spin-phonon interactions and the role of the translational diffusion of the $f$-electrons.
%
% model description
%

As a first step we find the renormalized relaxation rate of the transverse (relative to the external static magnetic field) magnetic moment of the Kondo ions toward the conduction electrons which remain in a thermodynamical equilibrium state (this is the Korringa relaxation rate $\Gamma_{ss}$). For an isotropic system with $g_\bot=g_\parallel=g$ the result in second order in $\mathcal{H}_{s\sigma }$ is well known: in terms of Eq. (\ref{Hssigma}) $\Gamma _{ss}  = \frac{\pi }{\hbar }\left( {g\rho J} \right)^2 k_{\rm B} T$. However, in the case of an antiferromagnetic coupling, $J > 0$, second order at low temperatures is not sufficient, and we have to improve perturbative calculations. This can be done by the renormalization of the coupling constant in the spirit of the AndersonÕs ``Poor Man's'' scaling.\cite{anderson70a} We provide our results for the static magnetic field oriented perpendicular to the crystal symmetry axis only, since the most detailed experiments with YbRh$_{2}$Si$_{2}$ were performed for this orientation and, moreover, $g_\parallel$ data of YbRh$_{2}$Si$_{2}$ were not accessible with the available experimental equipment due to $g_\parallel < 0.2$. The renormalized Korringa relaxation rate was found for $k_{\rm B} T > B$ using the method of functional variational derivatives: 
\begin{align}
\Gamma _{\rm ss}  =& \frac{\pi }{\hbar }(\rho J)^2 \left( {g_ \bot ^2  - g_\parallel ^2 } \right)k_{\rm B} T\left( {\cot ^2 \varphi  + \frac{3}{4}} \right);\label{Gss}\\
\varphi  =  & \rho J\sqrt {g_ \bot ^2  - g_\parallel ^2 } \ln \left( {T/T_{\rm GK} } \right). \nonumber
\end{align}
The parameter $\varphi$ reflects the logarithmic behaviour at temperatures close to the characteristic value $T_{\rm GK}$. For the Kramers ground state we have found
\begin{equation}
T_{\rm GK}  = W\!\exp\!\!\left[ { - \frac{1}{{\rho J\sqrt {g_ \bot ^2  - g_\parallel ^2 } }}{\mathop{\rm arc}\nolimits}\cot\!\!\left(\!{\frac{{g_\parallel  }}{{\sqrt {g_ \bot ^2  - g_\parallel ^2 } }}}\!\right)} \!\right] \label{TGK}
\end{equation}
with $W$ as the band width of the conduction electrons. One can see that in the isotropic case the standard result $T_{\rm K}  = W\exp \left[  - \frac{1}{\rho Jg} \right]$ can be obtained asymtotically for $g_ \bot =g_\parallel $ from Eq. (\ref{TGK}). As expected, the Korringa relaxation rate is logarithmically divergent upon lowering the temperature to $T_{\rm GK}(\varphi\rightarrow0): \Gamma _{\rm ss}  \propto 1/\ln ^2 \left( {T/T_{\rm GK} } \right)$. The Overhauser relaxation rate $\Gamma _{\sigma \sigma } $ (magnetic moment of the conduction electrons relaxes toward the Yb$^{3+}$ spin system, being in the equilibrium with a thermal bath) can be found from the relation 
\begin{equation}
\Gamma _{\rm ss} /\Gamma _{\sigma \sigma }  = 2\rho k_{\rm B} (T + \theta_\bot )g_\sigma  /g_ \bot \;,\;\theta_\bot  = \frac{1}{{4k_{\rm B} }}\sum\limits_j {I_{ij}^ \bot  } .\label{Gsssigsig}
\end{equation}
Here $\theta_\bot$ is the Weiss temperature, which originates from the RKKY interaction in the molecular field approximation. Evidently, $\Gamma _{\sigma \sigma }$ is also divergent as $\Gamma _{\sigma \sigma }\propto 1/\ln ^2 \left( {T/T_{\rm GK} } \right)$. At first glance, these results confirm the commonly accepted belief that the ESR linewidth of Kondo ions (as well as conduction electrons) is expected to be too large for its detection. However, an equilibrium-state approximation for the conduction-electron-spin system is not valid to study the ESR response of the samples with a high concentration of Kondo ions \cite{barnes81a} which is especially the case for the Kondo lattice. Instead one has to derive coupled kinetic equations for both magnetic moments of Kondo ions and conduction electrons.

The kinetic equations of motion for the transverse magnetizations of localized moments and conduction electrons are coupled by two additional kinetic coefficients $\Gamma _{\rm s\sigma}$ and $\Gamma _{\sigma \rm s}$, respectively. For a correct analysis of a stationary solution one has to introduce, also, relaxation rates of the Kondo spin system and conduction electrons toward the thermal bath (``lattice'') $\Gamma _{ \rm sL}$ and $\Gamma _{\sigma \rm L}$, respectively. For the anisotropic system such equations were derived in second order of the Kondo interaction by the methods of nonequilibrium statistical operator and Green functions \cite{kochelaev04a}. The renormalized kinetic coefficients $\Gamma _{\rm s\sigma}$ and $\Gamma _{\sigma \rm s}$  were found for $k_{\rm B} T > B$ to be:  
\begin{equation}
\Gamma _{\sigma \rm s}  = \frac{\pi }{{4\hbar }}(\rho J)^2 \left( {g_ \bot ^2  - g_\parallel ^2 } \right)k_{\rm B} T\frac{1}{{\sin ^2 \left( {\varphi /2} \right)}} \label{Gsigmas}
\end{equation}
$\Gamma _{\rm s\sigma}$ can be found from a relation similar to Eq. (\ref{Gsssigsig}). One sees that both coefficients are also divergent in the same way (via the parameter $\varphi$) as the Korringa and Overhauser relaxation rates. To study the ESR response of the total system, one has to look for the poles of a solution of coupled equations after their time Fourier transform. As a result we obtain two complex frequencies: their real parts give resonant frequencies, their imaginary parts give the corresponding relaxation rates. We are interested in a solution close to the Kondo-ion resonance frequency.
%
%	 Òanisotropic Kondo bottleneckÓ regime. 
%

The coupling between two systems is especially important if the relaxation rate of conduction electrons toward the Kondo ions is much faster than to the lattice and the resonant frequencies are close to one another (``bottleneck'' regime):
$
\Gamma _{\sigma \sigma }  \gg \Gamma _{\sigma \rm L} ,|\omega _{\rm s}  - \omega _\sigma  |
$.
It is well known that in the case of an isotropic system and equal Larmor frequencies, the ESR linewidth in the bottleneck regime is greatly narrowed due to conservation of the total magnetic moment (its operator commutes with the isotropic Kondo interaction and the latter disappears from the effective relaxation rate). The same situation remains, if one takes into account the Kondo effect \cite{fazleev92a}. In the opposite case of a strongly anisotropic Kondo interaction the results obtained in second order do not show any sufficient narrowing of the ESR linewidth in the bottleneck regime \cite{kochelaev04a}. However, the renormalization of all kinetic coefficients makes the situation completely different. The relaxation rate of the collective mode with a frequency 
close to the Kondo-ion resonance now follows for $T > T_{\rm GK}$:
\begin{align}
\Gamma _{\rm coll}  = \Gamma _{\rm sL}  + \Gamma _{\rm ss}^{\rm eff} & + \Gamma _{\sigma\rm L}^{\rm eff}  + B^2F(T); \label{Geff1}\\
\Gamma _{\rm ss}^{\rm eff}  = \Gamma _{\rm ss}  - \Gamma _{\rm s\sigma } \Gamma _{\sigma \rm s} /\Gamma _{\sigma \sigma },&%
\hspace{2ex}
\Gamma _{\sigma \rm L}^{\rm eff}  = \Gamma _{\sigma \rm L} \left( {\Gamma _{\rm s\sigma } \Gamma _{\sigma \rm s} /\Gamma _{\sigma \sigma }^2 } \right) \nonumber
\end{align}
The simplified expressions for an effective Korringa relaxation rate $\Gamma _{\rm ss}^{\rm eff}$ and an effective relaxation rate of conduction electrons to the lattice $\Gamma _{\sigma \rm L}^{\rm eff}$ in the case $\varphi < $1 are:
\begin{align}
\Gamma _{\rm ss}^{\rm eff} (\varphi  < 1) = & \frac{\pi }{8}\left( {\rho J} \right)^4 \left( {g_ \bot ^2  - g_\parallel ^2 } \right)^2 k_{\rm B} T\ln ^2 \frac{T}{{T_{\rm GK} }},\nonumber\\
\Gamma _{\sigma \rm L}^{\rm eff} (\varphi  < 1) = & 2\rho k_{\rm B} (T+\theta_\bot)\Gamma _{\sigma \rm L} g_\sigma/g_ \bot.
\label{Geff2}
\end{align}
$B^2F(T)$ describes the external magnetic field dependence which is a consequence of a partial ÒopeningÓ of the bottleneck at large magnetic fields due to different $g$-factors of the Kondo ions and conduction electrons. The coefficient $F(T)$ is a rather extended expression containing the temperature dependence. 
It is important to note that, instead of being divergent, the effective Korringa relaxation rate $\Gamma _{\rm ss}^{\rm eff}$ is greatly reduced and goes to zero at $T\rightarrow T_{\rm GK}$ ! This is a remarkable result: although the total magnetic moment does not commute with the strongly anisotropic Kondo interaction at all, the divergent parts of all the kinetic coefficients cancel each other in the collective spin mode due to the existence of the common energy scale $T_{\rm GK}$, regulating their temperature dependence at $T\rightarrow T_{\rm GK}$. The effective relaxation rate of conduction electrons to the lattice $\Gamma _{\sigma \rm L}^{\rm eff}$ is also greatly reduced, becoming proportional to temperature and mimicking the usual Korringa relaxation rate. These results allow one to conclude that the main reason of the observability of ESR in a Kondo lattice is a formation of a collective spin mode in the bottleneck regime and in the presence of a Kondo effect. Another important ingredient -- the short-range ferromagnetic fluctuations due to the RKKY interactions will be discussed below.
%
%motional narrowing
%

Now we have to consider the broadening of the ESR linewidth which is represented by the kinetic coefficient $\Gamma _{\rm sL}$. An obvious contribution comes from the distribution of effective local magnetic fields due to spin-spin interactions of the Yb-ions and a variation of the $g$-factors due to defects of the lattice. In particular, the usual magnetic dipole-dipole interactions yield approximately $\Delta H_{\rm loc}\simeq700$~Oe, while the observed ESR linewidth in YbRh$_{2}$Si$_{2}$ at the X and Q bands is $\Delta H_{\rm ESR}\simeq 200$~Oe at $T = 5$~K. The contribution from the RKKY  interactions, which become highly anisotropic after projection onto the Kramers ground state, is expected to be much larger. Therefore, it is evident that some narrowing mechanism for these type of contributions should exist. 

It is well known that the broadening of the ESR line by the distribution of local fields can be reduced in the bottleneck regime due to fast reorientation of the Kondo ion moment caused by the Korringa relaxation \cite{barnes81a}. However, we expect that in the Kondo lattice the motional narrowing due to translational diffusion of quasilocalized $f$-electrons in the NFL state can be much more effective. An elementary step of the diffusion is a jump of an $f$-electron (or a hole in the closed $f$-shell) from one Yb site to the next nearest site. At the same time this jump can be easily done, if the local field, created by the RKKY interaction, is the same. This can happen only, if the RKKY interaction for the nearest Yb ions is \textit{ferromagnetic}. It is worth to mention that the importance of ferromagnetic fluctuations for the ESR observability at $T < T_{\rm K}$ in Kondo lattice with heavy fermions was discussed recently in a different context \cite{abrahams08a,schlottmann09a}.

An additional broadening of the ESR linewidth can appear due to the spin-phonon interaction of the Kondo ions. The main contribution comes from the two-phonon Raman and Orbach processes at temperatures above a few K. The Orbach process via the excited energy level at a given crystal field splitting $\Delta$ yields the temperature dependence 
$ \Gamma _{\rm sL}^{\rm Orbach}  \propto \left[ \exp ( \Delta /k_{\rm B} T ) - 1\right]^{-1} .$
The same temperature dependence is provided by the scattering of conduction electrons via the excited energy level.

Concerning the ESR resonance frequency of the collective mode, the situation is somewhat different. For a single Kondo ion it is well known that besides the usual Knight shift of the ESR resonance frequency, the Kondo effect results in a divergent logarithmic term. The same happens with the resonance frequency of the conduction electrons. We have found that all divergent parts of the ESR resonance frequency cancel each other in the collective mode similar to the relaxation rate described above.  However, the RKKY interaction provides an additional local field at the Yb-ion and the Weiss constant $\theta_\bot$ in the spin susceptibility. In the molecular field approximation involving both the Kondo and RKKY interactions, $\theta_\bot$ becomes also subject of the Kondo renormalization. As a result, the ESR resonance frequency contains the divergent logarithmic term even for the collective spin mode. For the corresponding effective $g$-factor $g_\perp^{\rm eff}$ we obtain the following relation (with the magnetic field perpendicular to the symmetry axis; less important non-divergent terms are omitted): 
\begin{align}
\frac{g_ \bot}{g_\bot^{\rm eff}} =&1+\frac{\theta_\bot }{T}\bigg\{1 + \rho J \bigg[ g_\parallel   + \sqrt {g_ \bot ^2  - g_\parallel ^2 }\;\times \nonumber\\
& \times \left({\frac{1}{2}{\mathop{\rm arc}\nolimits} \cot \left( {g_\parallel  /\sqrt {g_ \bot ^2  - g_\parallel ^2 } } \right) - \cot \varphi}\right) \bigg] \bigg\}.
\label{Geff3}
\end{align}
%
% model application to data
%

\begin{figure}
\begin{center}
\includegraphics[width=0.75\columnwidth,clip]{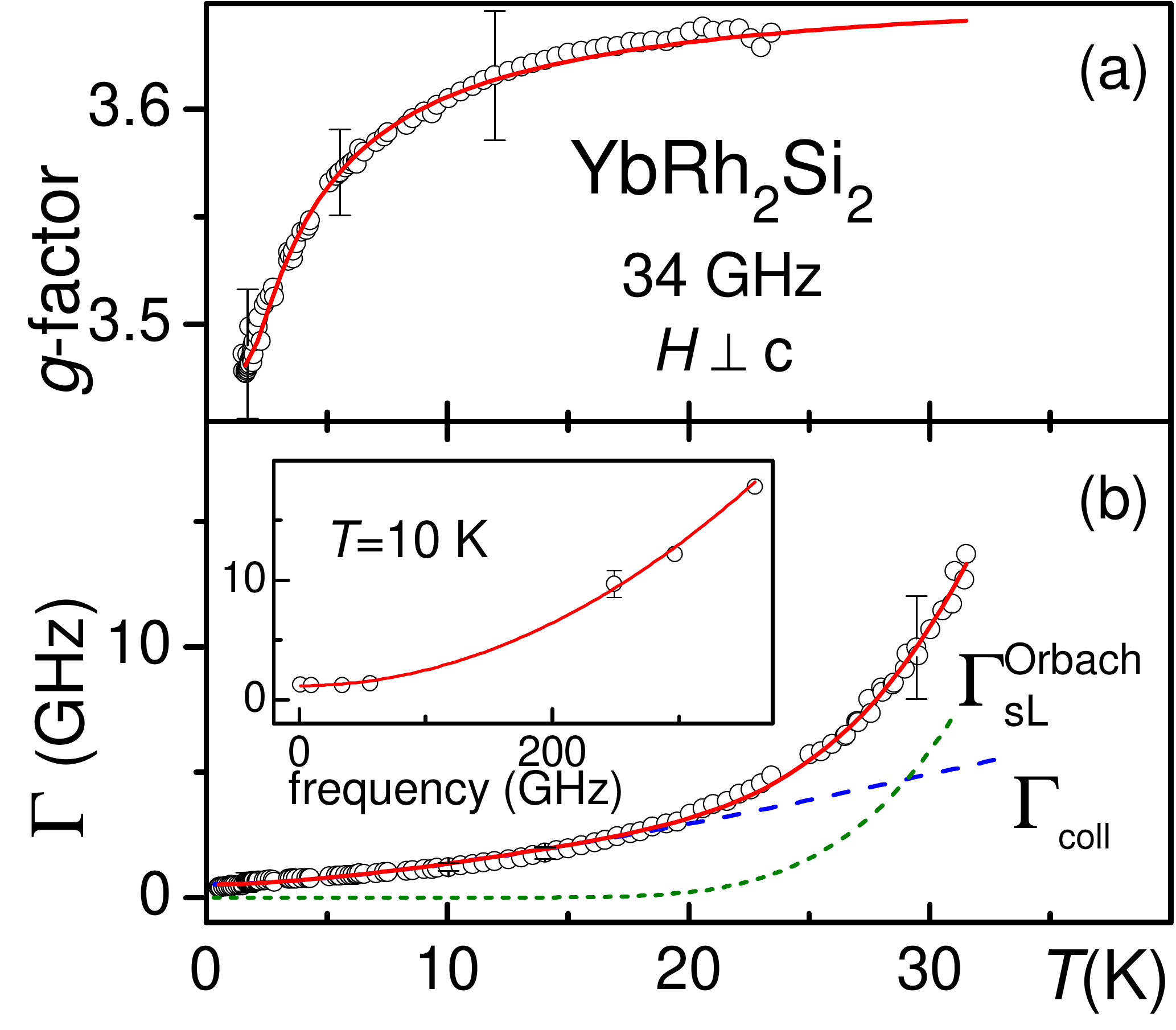}
\caption{Temperature dependence of Q-band (a) $g$-factor and (b) ESR relaxation rate $\Gamma$ of YbRh$_{2}$Si$_{2}$ (open circles). Solid lines denote data fitting: $g_ \bot ^{\rm eff}$ (Eq. (\ref{Geff3})) and $\Gamma _{\rm theor}$ (with two contributions as indicated).
Inset: Frequency dependence of $\Gamma _{\rm theor}$ (solid line) fitted to $\Gamma(T=10\;\text{K})$ (data from Refs.\onlinecite{sichelschmidt03a,schaufuss09a}).}\label{Fig1}
\end{center}
\end{figure}

Next, we compare our theory with experimental results. At first, we use the divergent logarithmic term in $g_ \bot ^{\rm eff}$ to reveal the characteristic temperature $T_{\rm GK}$. Starting values of $g$-factors were taken from the crystal field consideration ($g_ \bot = 3.66$) \cite{kutuzov08a}, the density of states can be related to the band width of the conduction electrons as $\rho = 1/W$.  The result of the fitting is given in Fig. \ref{Fig1}a with $\rho J = 0.05, \theta_\bot= 0.18$~K and $T_{\rm GK} = 0.36$~K; the latter is by two orders of magnitude smaller than the Kondo temperature $T_{\rm K}$ derived thermodynamically \cite{trovarelli00a} and by transport measurements \cite{kohler08a}.
Although the revealed value $T_{\rm GK} = 0.36$~K could be rather approximate, we used it to fit the temperature and frequency dependencies of  the ESR linewidth with help of the Eqs. (\ref{Geff1},\ref{Geff2}) in order to see whether our theory is selfconsistent:
\begin{equation}
\Gamma _{\rm theor}  = \Gamma _{\rm coll}  + \Gamma _{\rm sL}^{\rm Orbach} + Const. 
\label{Gammatheor}
\end{equation}
Here $Const.$ represents the local field distribution which is greatly reduced by the motional narrowing mechanism as discussed above. The results are given in Fig. \ref{Fig1}b. The fitting of the temperature dependence of the ESR linewidth gave $\rho J = 0.048$ and $\Delta = 198$~K. The latter coincides with the first excited energy level of the Yb-ion\cite{stockert06a}, confirming that the Orbach processes dominate in the spin-phonon relaxation (our estimation of the spin relaxation with the conduction electrons via excited level gave a significantly smaller contribution). It can be assumed that the effective relaxation rate $\Gamma _{\sigma\rm L}^{\rm eff}$ of conduction electrons to the lattice is negligible.

From the inset of Fig. \ref{Fig1}b one can see that $\Gamma _{\rm theor}$ increases roughly as $\nu^2$, in full accord with the experimental observation. From the frequency dependence of the ESR linewidth we obtain $\rho J = 0.07$; different fittings resulted in only slight variations of $\rho J$. Crude approximations like $\rho=1/W$ may already explain such variations. Having the value $\rho J$, we can now estimate the Korringa relaxation rate without the bottleneck regime. According to Eq. (\ref{Gss}) the value $\rho J = 0.05$ yields $\Gamma_{\rm ss} = 51$~GHz at 5~K. This would leave no chance to observe the ESR signal, neither at X- or Q-bands nor at higher frequencies without formation of the collective spin mode. It is remarkable that the Kondo effect, being responsible for a suppression of the ESR signal on paramagnetic impurities in a metal at low temperatures $T < T_{\rm K}$ , crucially supports it in the Kondo lattice due to the formation of the collective spin mode with a dramatic narrowing of the ESR linewidth, even in the case of a strongly anisotropic Kondo interaction. Recently, additional experimental arguments in favor of the collective spin mode were given in Ref. \onlinecite{duque09a}. We should mention that the relation between the thermodynamical Kondo temperature $T_{\rm K}$ and our characteristic temperature for the ground Kramers doublet $T_{\rm GK}$ remains an open question in our consideration. Probably, it could be related to the Kondo resonance narrowing considered quite recently.\cite{nevidomskyy09a}\\
\indent In conclusion, we have revealed new features in the properties of the Kondo lattice with heavy fermions and have found an explanation of the ESR observability in YbRh$_{2}$Si$_{2}$, that was in the last years a subject of sharp discussions in the literature.\\
\indent This work was supported by the Volkswagen Foundation (I/82203); AMS was partially supported by the cooperative grant Y4-P-07-15 from the CRDF BRHE program, ASK and BIK were partially supported by the RMES program ÒDevelopment of scientific potential of higher schoolÓ (2.1.1/2985). 
%
%
%\bibliography{JoergBib}

\begin{thebibliography}{21}
\expandafter\ifx\csname natexlab\endcsname\relax\def\natexlab#1{#1}\fi
\expandafter\ifx\csname bibnamefont\endcsname\relax
  \def\bibnamefont#1{#1}\fi
\expandafter\ifx\csname bibfnamefont\endcsname\relax
  \def\bibfnamefont#1{#1}\fi
\expandafter\ifx\csname citenamefont\endcsname\relax
  \def\citenamefont#1{#1}\fi
\expandafter\ifx\csname url\endcsname\relax
  \def\url#1{\texttt{#1}}\fi
\expandafter\ifx\csname urlprefix\endcsname\relax\def\urlprefix{URL }\fi
\providecommand{\bibinfo}[2]{#2}
\providecommand{\eprint}[2][]{\url{#2}}

\bibitem[{\citenamefont{Sichelschmidt et~al.}(2003)\citenamefont{Sichelschmidt,
  Ivanshin, Ferstl, Geibel, and Steglich}}]{sichelschmidt03a}
\bibinfo{author}{\bibfnamefont{J.}~\bibnamefont{Sichelschmidt}},
  \bibinfo{author}{\bibfnamefont{V.~A.}~\bibnamefont{Ivanshin}},
  \bibinfo{author}{\bibfnamefont{J.}~\bibnamefont{Ferstl}},
  \bibinfo{author}{\bibfnamefont{C.}~\bibnamefont{Geibel}}, \bibnamefont{and}
  \bibinfo{author}{\bibfnamefont{F.}~\bibnamefont{Steglich}},
  \bibinfo{journal}{Phys.\ Rev.\ Lett.} \textbf{\bibinfo{volume}{91}},
  \bibinfo{pages}{156401} (\bibinfo{year}{2003}).

\bibitem[{\citenamefont{Wykhoff et~al.}(2007)\citenamefont{Wykhoff,
  Sichelschmidt, Lapertot, Knebel, Flouquet, Fazlishanov, Krug~von Nidda,
  Krellner, Geibel, and Steglich}}]{wykhoff07b}
\bibinfo{author}{\bibfnamefont{J.}~\bibnamefont{Wykhoff}},
  \bibinfo{author}{\bibfnamefont{J.}~\bibnamefont{Sichelschmidt}},
  \bibinfo{author}{\bibfnamefont{G.}~\bibnamefont{Lapertot}},
  \bibinfo{author}{\bibfnamefont{G.}~\bibnamefont{Knebel}},
  \bibinfo{author}{\bibfnamefont{J.}~\bibnamefont{Flouquet}},
  \bibinfo{author}{\bibfnamefont{I.~I.} \bibnamefont{Fazlishanov}},
  \bibinfo{author}{\bibfnamefont{H.-A.} \bibnamefont{Krug~von Nidda}},
  \bibinfo{author}{\bibfnamefont{C.}~\bibnamefont{Krellner}},
  \bibinfo{author}{\bibfnamefont{C.}~\bibnamefont{Geibel}}, \bibnamefont{and}
  \bibinfo{author}{\bibfnamefont{F.}~\bibnamefont{Steglich}},
  \bibinfo{journal}{Science Techn. Adv. Mat.} \textbf{\bibinfo{volume}{8}},
  \bibinfo{pages}{389} (\bibinfo{year}{2007}).

\bibitem[{\citenamefont{Sichelschmidt et~al.}(2007)\citenamefont{Sichelschmidt,
  Wykhoff, {H.-A. Krug von Nidda}, Fazlishanov, Hossain, Krellner, Geibel, and
  Steglich}}]{sichelschmidt07a}
\bibinfo{author}{\bibfnamefont{J.}~\bibnamefont{Sichelschmidt}},
  \bibinfo{author}{\bibfnamefont{J.}~\bibnamefont{Wykhoff}},
  \bibinfo{author}{\bibnamefont{{H.-A. Krug von Nidda}}},
  \bibinfo{author}{\bibfnamefont{I.~I.} \bibnamefont{Fazlishanov}},
  \bibinfo{author}{\bibfnamefont{Z.}~\bibnamefont{Hossain}},
  \bibinfo{author}{\bibfnamefont{C.}~\bibnamefont{Krellner}},
  \bibinfo{author}{\bibfnamefont{C.}~\bibnamefont{Geibel}}, \bibnamefont{and}
  \bibinfo{author}{\bibfnamefont{F.}~\bibnamefont{Steglich}},
  \bibinfo{journal}{J. Phys. Cond. Mat.} \textbf{\bibinfo{volume}{19}},
  \bibinfo{pages}{016211} (\bibinfo{year}{2007}).

\bibitem[{\citenamefont{Schaufuss et~al.}(2009)\citenamefont{Schaufuss, Kataev,
  Zvyagin, Buchner, Sichelschmidt, Wykhoff, Krellner, Geibel, and
  Steglich}}]{schaufuss09a}
\bibinfo{author}{\bibfnamefont{U.}~\bibnamefont{Schaufuss}},
  \bibinfo{author}{\bibfnamefont{V.}~\bibnamefont{Kataev}},
  \bibinfo{author}{\bibfnamefont{A.~A.} \bibnamefont{Zvyagin}},
  \bibinfo{author}{\bibfnamefont{B.}~\bibnamefont{Buchner}},
  \bibinfo{author}{\bibfnamefont{J.}~\bibnamefont{Sichelschmidt}},
  \bibinfo{author}{\bibfnamefont{J.}~\bibnamefont{Wykhoff}},
  \bibinfo{author}{\bibfnamefont{C.}~\bibnamefont{Krellner}},
  \bibinfo{author}{\bibfnamefont{C.}~\bibnamefont{Geibel}}, \bibnamefont{and}
  \bibinfo{author}{\bibfnamefont{F.}~\bibnamefont{Steglich}},
  \bibinfo{journal}{Phys. Rev. Lett.} \textbf{\bibinfo{volume}{102}},
  \bibinfo{eid}{076405} (\bibinfo{year}{2009}).

\bibitem[{\citenamefont{Si et~al.}(2001)\citenamefont{Si, Rabello, Ingersent,
  and Smith}}]{si01a}
\bibinfo{author}{\bibfnamefont{Q.}~\bibnamefont{Si}},
  \bibinfo{author}{\bibfnamefont{S.}~\bibnamefont{Rabello}},
  \bibinfo{author}{\bibfnamefont{K.}~\bibnamefont{Ingersent}},
  \bibnamefont{and} \bibinfo{author}{\bibfnamefont{L.}~\bibnamefont{Smith}},
  \bibinfo{journal}{Nature} \textbf{\bibinfo{volume}{413}},
  \bibinfo{pages}{804} (\bibinfo{year}{2001}).

\bibitem[{\citenamefont{Gegenwart et~al.}(2008)\citenamefont{Gegenwart, Si, and
  Steglich}}]{gegenwart08a}
\bibinfo{author}{\bibfnamefont{P.}~\bibnamefont{Gegenwart}},
  \bibinfo{author}{\bibfnamefont{Q.}~\bibnamefont{Si}}, \bibnamefont{and}
  \bibinfo{author}{\bibfnamefont{F.}~\bibnamefont{Steglich}},
  \bibinfo{journal}{Nature Phys.} \textbf{\bibinfo{volume}{4}},
  \bibinfo{pages}{186} (\bibinfo{year}{2008}).

\bibitem[{\citenamefont{Kutuzov et~al.}(2008)\citenamefont{Kutuzov, Skvortsova,
  Belov, Sichelschmidt, Wykhoff, Eremin, Krellner, Geibel, and
  Kochelaev}}]{kutuzov08a}
\bibinfo{author}{\bibfnamefont{A.}~\bibnamefont{Kutuzov}},
  \bibinfo{author}{\bibfnamefont{A.}~\bibnamefont{Skvortsova}},
  \bibinfo{author}{\bibfnamefont{S.}~\bibnamefont{Belov}},
  \bibinfo{author}{\bibfnamefont{J.}~\bibnamefont{Sichelschmidt}},
  \bibinfo{author}{\bibfnamefont{J.}~\bibnamefont{Wykhoff}},
  \bibinfo{author}{\bibfnamefont{I.}~\bibnamefont{Eremin}},
  \bibinfo{author}{\bibfnamefont{C.}~\bibnamefont{Krellner}},
  \bibinfo{author}{\bibfnamefont{C.}~\bibnamefont{Geibel}}, \bibnamefont{and}
  \bibinfo{author}{\bibfnamefont{B.}~\bibnamefont{Kochelaev}},
  \bibinfo{journal}{J. Phys. Cond. Mat.} \textbf{\bibinfo{volume}{20}},
  \bibinfo{pages}{455208} (\bibinfo{year}{2008}).

\bibitem[{\citenamefont{Gegenwart et~al.}(2005)\citenamefont{Gegenwart,
  Custers, Tokiwa, Geibel, and Steglich}}]{gegenwart05a}
\bibinfo{author}{\bibfnamefont{P.}~\bibnamefont{Gegenwart}},
  \bibinfo{author}{\bibfnamefont{J.}~\bibnamefont{Custers}},
  \bibinfo{author}{\bibfnamefont{Y.}~\bibnamefont{Tokiwa}},
  \bibinfo{author}{\bibfnamefont{C.}~\bibnamefont{Geibel}}, \bibnamefont{and}
  \bibinfo{author}{\bibfnamefont{F.}~\bibnamefont{Steglich}},
  \bibinfo{journal}{Phys. Rev. Lett.} \textbf{\bibinfo{volume}{94}},
  \bibinfo{pages}{076402} (\bibinfo{year}{2005}).

\bibitem[{\citenamefont{Hossain et~al.}(2005)\citenamefont{Hossain, Geibel,
  Weickert, Radu, Tokiwa, Jeevan, Gegenwart, and Steglich}}]{hossain05a}
\bibinfo{author}{\bibfnamefont{Z.}~\bibnamefont{Hossain}},
  \bibinfo{author}{\bibfnamefont{C.}~\bibnamefont{Geibel}},
  \bibinfo{author}{\bibfnamefont{F.}~\bibnamefont{Weickert}},
  \bibinfo{author}{\bibfnamefont{T.}~\bibnamefont{Radu}},
  \bibinfo{author}{\bibfnamefont{Y.}~\bibnamefont{Tokiwa}},
  \bibinfo{author}{\bibfnamefont{H.}~\bibnamefont{Jeevan}},
  \bibinfo{author}{\bibfnamefont{P.}~\bibnamefont{Gegenwart}},
  \bibnamefont{and} \bibinfo{author}{\bibfnamefont{F.}~\bibnamefont{Steglich}},
  \bibinfo{journal}{Phys.\ Rev.\ B} \textbf{\bibinfo{volume}{72}},
  \bibinfo{pages}{094411} (\bibinfo{year}{2005}).

\bibitem[{\citenamefont{Stockert et~al.}(2006)\citenamefont{Stockert, Koza,
  Ferstl, Murani, Geibel, and Steglich}}]{stockert06a}
\bibinfo{author}{\bibfnamefont{O.}~\bibnamefont{Stockert}},
  \bibinfo{author}{\bibfnamefont{M.~M.} \bibnamefont{Koza}},
  \bibinfo{author}{\bibfnamefont{J.}~\bibnamefont{Ferstl}},
  \bibinfo{author}{\bibfnamefont{A.~P.} \bibnamefont{Murani}},
  \bibinfo{author}{\bibfnamefont{C.}~\bibnamefont{Geibel}}, \bibnamefont{and}
  \bibinfo{author}{\bibfnamefont{F.}~\bibnamefont{Steglich}},
  \bibinfo{journal}{Physica B} \textbf{\bibinfo{volume}{378-380}},
  \bibinfo{pages}{157} (\bibinfo{year}{2006}).

\bibitem[{\citenamefont{Anderson}(1970)}]{anderson70a}
\bibinfo{author}{\bibfnamefont{P.~W.} \bibnamefont{Anderson}},
  \bibinfo{journal}{J. Phys. C: Solid State Phys.}
  \textbf{\bibinfo{volume}{3}}, \bibinfo{pages}{2436} (\bibinfo{year}{1970}).

\bibitem[{\citenamefont{Barnes}(1981)}]{barnes81a}
\bibinfo{author}{\bibfnamefont{S.~E.} \bibnamefont{Barnes}},
  \bibinfo{journal}{Adv. Phys.} \textbf{\bibinfo{volume}{30}},
  \bibinfo{pages}{801} (\bibinfo{year}{1981}).

\bibitem[{\citenamefont{Kochelaev and Safina}(2004)}]{kochelaev04a}
\bibinfo{author}{\bibfnamefont{B.~I.} \bibnamefont{Kochelaev}}
  \bibnamefont{and} \bibinfo{author}{\bibfnamefont{A.~M.}
  \bibnamefont{Safina}}, \bibinfo{journal}{Phys. Sol. State}
  \textbf{\bibinfo{volume}{46}}, \bibinfo{pages}{226} (\bibinfo{year}{2004}).

\bibitem[{\citenamefont{Fazleev et~al.}(1992)\citenamefont{Fazleev, Mironov,
  and Fry}}]{fazleev92a}
\bibinfo{author}{\bibfnamefont{N.~G.} \bibnamefont{Fazleev}},
  \bibinfo{author}{\bibfnamefont{G.~I.} \bibnamefont{Mironov}},
  \bibnamefont{and} \bibinfo{author}{\bibfnamefont{J.~L.} \bibnamefont{Fry}},
  \bibinfo{journal}{J. Magn. Mag. Mat.} \textbf{\bibinfo{volume}{108}},
  \bibinfo{pages}{123} (\bibinfo{year}{1992}).

\bibitem[{\citenamefont{Abrahams and W\"{o}lfle}(2008)}]{abrahams08a}
\bibinfo{author}{\bibfnamefont{E.}~\bibnamefont{Abrahams}} \bibnamefont{and}
  \bibinfo{author}{\bibfnamefont{P.}~\bibnamefont{W\"{o}lfle}},
  \bibinfo{journal}{Phys. Rev. B} \textbf{\bibinfo{volume}{78}},
  \bibinfo{eid}{104423} (\bibinfo{year}{2008}).

\bibitem[{\citenamefont{Schlottmann}(2009)}]{schlottmann09a}
\bibinfo{author}{\bibfnamefont{P.}~\bibnamefont{Schlottmann}},
  \bibinfo{journal}{Phys. Rev. B} \textbf{\bibinfo{volume}{79}},
  \bibinfo{eid}{045104} (\bibinfo{year}{2009}).

\bibitem[{\citenamefont{Abragam and Bleaney}(1970)}]{abragam70a}
\bibinfo{author}{\bibfnamefont{A.}~\bibnamefont{Abragam}} \bibnamefont{and}
  \bibinfo{author}{\bibfnamefont{B.}~\bibnamefont{Bleaney}},
  \emph{\bibinfo{title}{Electron Paramagnetic Resonance of Transition Ions}}
  (\bibinfo{publisher}{Clarendon Press}, \bibinfo{address}{Oxford},
  \bibinfo{year}{1970}).

\bibitem[{\citenamefont{Trovarelli et~al.}(2000)\citenamefont{Trovarelli,
  Geibel, Mederle, Langhammer, Grosche, Gegenwart, Lang, Sparn, and
  Steglich}}]{trovarelli00a}
\bibinfo{author}{\bibfnamefont{O.}~\bibnamefont{Trovarelli}},
  \bibinfo{author}{\bibfnamefont{C.}~\bibnamefont{Geibel}},
  \bibinfo{author}{\bibfnamefont{S.}~\bibnamefont{Mederle}},
  \bibinfo{author}{\bibfnamefont{C.}~\bibnamefont{Langhammer}},
  \bibinfo{author}{\bibfnamefont{F.~M.} \bibnamefont{Grosche}},
  \bibinfo{author}{\bibfnamefont{P.}~\bibnamefont{Gegenwart}},
  \bibinfo{author}{\bibfnamefont{M.}~\bibnamefont{Lang}},
  \bibinfo{author}{\bibfnamefont{G.}~\bibnamefont{Sparn}}, \bibnamefont{and}
  \bibinfo{author}{\bibfnamefont{F.}~\bibnamefont{Steglich}},
  \bibinfo{journal}{Phys.\ Rev.\ Lett.} \textbf{\bibinfo{volume}{85}},
  \bibinfo{pages}{626} (\bibinfo{year}{2000}).

\bibitem[{\citenamefont{K\"ohler et~al.}(2008)\citenamefont{K\"ohler, Oeschler,
  Steglich, Maquilon, and Fisk}}]{kohler08a}
\bibinfo{author}{\bibfnamefont{U.}~\bibnamefont{K\"ohler}},
  \bibinfo{author}{\bibfnamefont{N.}~\bibnamefont{Oeschler}},
  \bibinfo{author}{\bibfnamefont{F.}~\bibnamefont{Steglich}},
  \bibinfo{author}{\bibfnamefont{S.}~\bibnamefont{Maquilon}}, \bibnamefont{and}
  \bibinfo{author}{\bibfnamefont{Z.}~\bibnamefont{Fisk}},
  \bibinfo{journal}{Phys. Rev. B} \textbf{\bibinfo{volume}{77}},
  \bibinfo{eid}{104412} (\bibinfo{year}{2008}).

\bibitem[{\citenamefont{Duque et~al.}(2009)\citenamefont{Duque, Bittar,
  Adriano, Giles, Holanda, Lora-Serrano, Pagliuso, Rettori, Perez, Hu
  et~al.}}]{duque09a}
\bibinfo{author}{\bibfnamefont{J.~G.~S.} \bibnamefont{Duque}},
  \bibinfo{author}{\bibfnamefont{E.~M.} \bibnamefont{Bittar}},
  \bibinfo{author}{\bibfnamefont{C.}~\bibnamefont{Adriano}},
  \bibinfo{author}{\bibfnamefont{C.}~\bibnamefont{Giles}},
  \bibinfo{author}{\bibfnamefont{L.~M.} \bibnamefont{Holanda}},
  \bibinfo{author}{\bibfnamefont{R.}~\bibnamefont{Lora-Serrano}},
  \bibinfo{author}{\bibfnamefont{P.~G.} \bibnamefont{Pagliuso}},
  \bibinfo{author}{\bibfnamefont{C.}~\bibnamefont{Rettori}},
  \bibinfo{author}{\bibfnamefont{C.~A.} \bibnamefont{Perez}},
  \bibinfo{author}{\bibfnamefont{R.}~\bibnamefont{Hu}}, \bibnamefont{et~al.},
  \bibinfo{journal}{Phys. Rev. B} \textbf{\bibinfo{volume}{79}},
  \bibinfo{eid}{035122} (\bibinfo{year}{2009}).

\bibitem[{\citenamefont{Nevidomskyy and Coleman}(2009)}]{nevidomskyy09a}
\bibinfo{author}{\bibfnamefont{A.~H.} \bibnamefont{Nevidomskyy}}
  \bibnamefont{and} \bibinfo{author}{\bibfnamefont{P.}~\bibnamefont{Coleman}},
  \bibinfo{journal}{arXiv} \textbf{\bibinfo{volume}{0906}},
  \bibinfo{pages}{4107v1} (\bibinfo{year}{2009}).

\end{thebibliography}

%
\end{document}